\documentstyle[graphicx]{elsart}

\begin{document}
\begin{frontmatter}
\title{General analysis of polarization phenomena in $ e^++e^-\rightarrow N+\bar N$ 
for axial parametrization of two--photon exchange}

\author{G. I. Gakh }
\thanks[ ]
{e-mail: \it gakh@kipt.kharkov.ua}
\address{
\it National Science Centre "Kharkov Institute of Physics and Technology"\\ 61108 Akademicheskaya 1, Kharkov,
Ukraine}
\author{E. Tomasi-Gustafsson}
\thanks[corr]{Corresponding author: etomasi@cea.fr}
\address{\it DAPNIA/SPhN, CEA/Saclay, 91191 Gif-sur-Yvette Cedex,
France }

\vspace{0.5cm}
\begin{abstract}
General properties of the differential cross section and of the polarization observables are derived for the $e^++e^-\rightarrow N+\bar N$ reaction, 
in presence of two--photon exchange. Polarization effects are investigated for longitudinally polarized electron and polarized antinucleon and/or polarized nucleon in the  final state. The single--spin asymmetry induced by the transverse polarization of the electron beam is also discussed.
\end{abstract}
\vspace{0.2cm}
%PACS: 12.20.-m, 13.40.-f, 13.60.-Hb, 13.88.+e

\end{frontmatter}
\maketitle
\section{Introduction}

The measurement of the nucleon electromagnetic form factors (FFs) in the 
space--like (SL) region of momentum transfer squared has a long history 
\cite{Ho62}. The 
electric and magnetic FFs were traditionally determined using the Rosenbluth separation  \cite{Ro50}, based on the measurement of the unpolarized differential cross section, for different angles, at the same momentum transfer. More recently, the polarization transfer method \cite{Re68}, which requires longitudinally polarized electron beam and the measurement of the recoil proton polarization, could be applied \cite{JG00}. It turned out that the ratio of the proton electric and magnetic FFs determined by the two methods, which are based on the same physics (one--photon exchange mechanism), lead to appreciably different results, and this difference is increasing 
with $Q^2$, the four-- momentum transfer squared \cite{Ar05}. One possible explanation for this discrepancy is the presence of a two--photon exchange (TPE) 
contribution to the elastic electron--nucleon scattering \cite{Gu03,Bl03,Ch04}.
Model independent properties of TPE in elastic electron--proton scattering have 
been derived in Ref. \cite{Re03}, where it was shown that the presence 
of such mechanism destroys the linearity of the Rosenbluth fit. No deviation from linearity is observed in the available data (in the limit of their precision) \cite{ETG05}. A previous study on deuteron target \cite{Re99} focussed already the attention to this problem, following a discrepancy observed in two precise  experiments on electron deuteron elastic scattering \cite{Al99,Ab99}.

Estimates of the TPE contribution to the elastic electron--deuteron
scattering were made in Refs. \cite{Gu73,Fr73} within the framework of the Glauber 
theory. It was shown \cite{Gu73} that this contribution decreases very slowly with 
momentum transfer squared $q^2$ and may dominate the cross section at high 
$q^2$ values. Since the TPE amplitude is essentially imaginary, the difference
between positron and electron scattering cross sections depends upon the small 
real part of the TPE amplitude \cite{Gu73}. Recoil polarization effects may be 
substantial, in the region where the one-- and two--photon exchange
contributions are comparable. If the TPE mechanism become sizeable, the 
straightforward  extraction  of FFs from the experimental data 
would no longer be possible \cite{Gu73}. It is known that double scattering
dominates in collisions of high--energy hadrons with deuterons at high $q^2$ 
values \cite{Fr73}, and in this paper it was predicted that the TPE effect in the elastic electron--deuteron scattering represents a 10 \%  effect at 
$q^2~\cong ~1.3$  GeV$^2$. At the same time the importance of the TPE mechanism was 
considered in Ref. \cite{Bo73}. The fact that the TPE mechanism, where the 
momentum transfer is shared between the two virtual photons, can become important 
with increasing $q^2$ value was already indicated more than thirty years ago 
\cite{Gu73,Fr73,Bo73}. 

Experimental investigation of nucleon FFs in the time--like (TL) region could shed new light on this problem and bring additional valuable information on the internal nucleon structure. In TL region the nucleon FFs can be accessed through the reactions 
$ e^++e^-\rightarrow N+\bar N$ or $\bar N+N\rightarrow  e^++e^-$.

The data on nucleon FFs in TL region are scarce and the determination of the individual electric and magnetic FFs is prevented by the statistics that it is 
possible to achieve \cite{An03}.  A  short review of the present status in this field 
of investigations can be found in Ref. \cite{Falco}. Unexpected results have been observed in the 
measurements of the nucleon FFs in the TL region (although the accuracy of the data is poor): the proton magnetic FF is smaller than the neutron one in the kinematical region covered by the experiments and the TL proton magnetic FF is considerably larger than the corresponding SL quantity. 

The neutron FFs were measured by the FENICE 
collaboration \cite{An98}, using the ADONE $e^+e^-$ collider in Frascati up to $q^2\approx$6 GeV$^2$. The proton FFs were measured in a more broad 
region of $q^2$ values. The differential cross section for $\bar p+p\rightarrow e^++e^-$ was measured by the Fermilab experiment E835  in the region of large $q^2$, up to $q^2$=18.22 GeV$^2$ \cite{An03}.

Recent and planned experiments focus on the TL region. Preliminary data from BaBar
\cite{Ba05}, using the radiative return, show surprising features as, for example, that the ratio 
$G_E/G_M$ is larger than one, in contradiction with polarization 
data from the SL region, and with previous TL data \cite{Ba94}. Moreover, TL FFs, still extracted under 
the hypothesis $|G_E|=|G_M|$, do not follow smoothly the asymptotic behavior predicted by QCD but 
show a few structures as a function of a $q^2$.
At the future antiproton facility, at GSI, a precise separation of FFs is planned  
\cite{PANDA} as well as polarization measurements \cite{PAX}. These ones will firstly allow to measure the 
relative phase of FFs, which are complex in the TL region. New measurements are also planned at Frascati, after an upgrade of $DA\Phi NE$ \cite{Fi03}, in particular for neutrons. The possibilty of accessing polarization observables is under study.

The reactions $ e^++e^-\rightarrow N+\bar N$ and $ N+\bar N\rightarrow e^++e^- $ are the crossing channel 
reactions of elastic electron--nucleon scattering. FFs describing the annihilation channel are assumed to be the analytical continuation of the SL ones. From crossing symmetry one expects that the reaction mechanisms are common for the scattering and the annihilation channels. This concerns, in particular, 
the problem of the TPE contribution. 

Theoretically the reaction $ e^++e^-\rightarrow N+\bar N$ was studied in a few papers. The dependence of the polarization of the emitted
baryons in $ e^++e^-\rightarrow B+\bar B $ on the polarization of 
colliding $ e^++e^-$-beams was firstly discussed in Ref. \cite{Du96}. All polarization effects for baryons with spin 1/2 were calculated, assuming one--photon exchange approximation. 
Numerical estimates of polarization effects were given for the nucleons, on the basis of two different versions of a VMD model. Polarization effects appear 
to be very sensitive to the choice of the nucleon FFs parametrization and are
rather large in absolute value. 

The existence of the T--odd single--spin asymmetry normal to the scattering 
plane requires a non--zero phase difference between the electric and magnetic 
FFs. The measurement of the polarization of one of the outgoing nucleons 
allows to determine the phase of the ratio $G_E/G_M$. In Ref. \cite{Br04} it 
was also shown that measurements of the proton polarization in $ e^++e^-\rightarrow
p+\bar p$ reaction  strongly discriminate between the analytic forms of different models 
suggested to describe the proton data in the SL region. In Ref. \cite{ETG05a} it was shown that this statement holds for other models, for double--spin polarization observables as well, and for the angular asymmetry, 
defined with respect to the cross section at 90$^0$.

We derive here the expressions for the differential cross section and various
polarization observables for the case when the matrix element of the reaction $ e^++e^-\rightarrow B+\bar B$
contains TPE contribution. The parametrization of the TPE term follows from 
the analytic continuation to the TL region of the approach used in the 
SL region in  Refs. \cite{Re03,Re04,Re04a}. 

The spin structure of the matrix element of the $ e^++e^-\rightarrow N+\bar N$ 
reaction is parametrized in terms of three independent complex functions depending
on two variables (energy and angle) rather than in terms of the standard nucleon FFs. 

This analysis follows the same steps as in Ref. \cite{Ga05}, where model independent properties 
of TPE were studied for the inverse reaction $p+\bar p\to e^++ e^-$ and polarization observables 
were derived using tensor parametrization for the TPE term. The purpose of this paper is to give 
explicit formulae for the observables, which will be useful for future experiments, as planned, 
for example, in Frascati \cite{Fi03} and 
Novosibirsk \cite{Ni05}. We analyze the  properties of the single-- and double--spin observables 
derived for the axial parametrization of the TPE term.
%%%%%%%%%%%%%%%%%%%%%%%%%%%%%%%%%%%
\section{Differential cross section}
%%%%%%%%%%%%%%%%%%%%%%%%%%%%%%%%%%%

The matrix element of the reaction $ e^++e^-\rightarrow N+\bar N$ can be obtained by analytic continuation of the matrix element for 
the elastic electron--nucleon scattering \cite{Re03,Fl63}, parametrizing the TPE contribution in axial form:
$${\cal M} = -\frac{e^2}{q^2}\biggl\{\bar u(-k_2)\gamma_{\mu }u(k_1)
\bar u(p_2)\biggl [F_{1N}(q^2, t)\gamma_{\mu }-\frac{1}{2m}F_{2N}(q^2, t)
\sigma_{\mu\nu }q_{\nu }\biggr ]u(-p_1)+ $$
\begin{equation}\label{eq:mat}
+\bar u(-k_2)\gamma_{\mu }\gamma_{5}u(k_1)\bar u(p_2)
\gamma_{\mu }\gamma_{5}u(-p_1)A_N(q^2, t)\biggr\}, 
\end{equation}
where $k_1~(k_2)$ and $p_1~(p_2)$ are the four--momenta of the electron 
(positron) and antinucleon (nucleon), respectively; $q=k_1+k_2=p_1+p_2,$ 
$ m $ is the nucleon mass.

The three complex amplitudes, $F_{1N,2N}(q^2, t)$ and $A_N(q^2, t),$ which 
generally are functions of two independent kinematical variables, $q^2$ 
and $t=(k_1-p_1)^2,$ fully describe the spin structure of the matrix element 
for the reaction $ e^++e^-\rightarrow N+\bar N$  - for any number of exchanged virtual photons.

This expression (\ref{eq:mat}) holds under assumption of the P--invariance of the
electromagnetic interaction and conservation of lepton helicity, which is 
correct for standard QED at the high energy, i.e., in zero electron mass limit. 
Note, however, that expression (\ref{eq:mat}) is one of the many equivalent representations
of the  $e^++e^-\rightarrow N+\bar N$ reaction matrix element.

In the Born (one--photon--exchange) approximation these amplitudes reduce to:
\begin{equation}\label{eq:eq3}
 F_{1N}^{Born}(q^2, t)=F_{1N}(q^2), \ 
 F_{2N}^{Born}(q^2, t)=F_{2N}(q^2), \
 A_N^{Born}(q^2, t)=0, \ 
\end{equation}
where $F_{1N}(q^2)$ and $F_{2N}(q^2)$ are the Dirac and Pauli nucleon
electromagnetic FFs, respectively, and they are complex functions 
of the variable $q^2.$ The complexity of FFs arises from the 
final--state strong interaction of the produced $N\bar N-$pair. In the 
following we use the standard magnetic $G_{MN}(q^2)$ and charge $G_{EN}(q^2)$ 
nucleon FFs which are related to FFs $F_{1N}(q^2)$ and $F_{2N}(q^2)$ as follows
\begin{equation}\label{eq:eq4}
 G_{MN}=F_{1N}+F_{2N}, \  G_{EN}=F_{1N}+\tau F_{2N}, \  
 \tau =\frac{q^2}{4m^2}, \  N=p, n.
\end{equation}
By analogy with these relations, let us introduce a linear combinations of the
$F_{1N,2N}(q^2, t)$ amplitudes which in the Born approximation correspond to the 
Sachs FFs $G_{MN}$ and $G_{EN}$:
\begin{eqnarray}
\tilde G_{MN}(q^2, t)&=&F_{1N}(q^2, t)+F_{2N}(q^2, t), \nonumber\\
\tilde G_{EN}(q^2, t)&=&F_{1N}(q^2, t)+\tau F_{2N}(q^2, t). 
\label{eq:eq5}
\end{eqnarray}
To separate the effects due to the Born and TPE contributions, let us single 
out the dominant contribution and define the following decompositions of the 
amplitudes 
\begin{eqnarray}
\tilde G_{MN}(q^2, t)&=&G_{MN}(q^2)+\Delta G_{MN}(q^2, t), \nonumber\\
\tilde G_{EN}(q^2, t)&=&G_{EN}(q^2)+\Delta G_{EN}(q^2, t). 
\label{eq:eq6}
\end{eqnarray}
$\Delta G_{MN}(q^2, t),$ $\Delta G_{EN}(q^2, t),$ and $A_N(q^2, t)$ are of the 
order of $\sim \alpha $, while $G_{MN}(q^2)$ and $G_{EN}(q^2)$ are of the order of $\sim \alpha^0.$ 
Since the terms $\Delta G_{MN},~\Delta G_{EN}$ and $A_N$ are small in 
comparison with the dominant ones, we neglect in the following the bilinear 
combinations of these terms.

We can rewrite the matrix element (\ref{eq:mat}) explicitating the vector and the axial terms as:
\begin{equation}\label{eq:eq7}
{\it M} = -\frac{e^2}{q^2}\biggl (j_{\mu }^{(v)}J_{\mu }^{(v)}+
j_{\mu }^{(a)}J_{\mu }^{(a)}\biggr ),
\end{equation}
where
$$j_{\mu }^{(v)}=\bar u(-k_2)\gamma_{\mu }u(k_1), \ 
J_{\mu }^{(v)}=\bar u(p_2)\biggl [\tilde G_{MN}\gamma_{\mu }-
\frac{\tilde G_{MN}-\tilde G_{EN}}{m(1-\tau )}P_{\mu }\biggr ]u(-p_1), $$
$$j_{\mu }^{(a)}=\bar u(-k_2)\gamma_{\mu }\gamma _5u(k_1), \ 
J_{\mu }^{(a)}=\bar u(p_2)\gamma_{\mu }\gamma _5u(-p_1)A_N, $$
and $P=(p_2-p_1)/2.$

Then the differential cross section of the reaction $ e^++e^-\rightarrow N+\bar N$ 
can be written as follows
in  the centre of mass system (CMS):
\begin{equation}\label{eq:eq8}
\frac{d\sigma}{d\Omega}=\frac{\alpha ^2\beta }{4q^6} 
\biggl [L_{\mu\nu }^{(v)}H_{\mu\nu }^{(v)}+
2Re(L_{\mu\nu }^{(i)}H_{\mu\nu }^{(i)})\biggr ], \
\end{equation}
$$L_{\mu\nu }^{(v)}=j_{\mu }^{(v)}j_{\nu}^{(v)*}, \ \ 
L_{\mu\nu }^{(i)}=j_{\mu }^{(a)}j_{\nu}^{(v)*},  \ \ 
H_{\mu\nu }^{(v)}=J_{\mu }^{(v)}J_{\nu}^{(v)*}, \ \
H_{\mu\nu }^{(i)}=J_{\mu }^{(a)}J_{\nu}^{(v)*}, $$
where $\beta $ is the nucleon velocity in the reaction CMS, $\beta =\sqrt{1-4m^2/q^2}$, and 
we neglected the terms proportional to $A_N^2.$

The leptonic tensors for the case of longitudinally polarized electrons have the forms 
\begin{eqnarray}
L_{\mu\nu }^{(v)}&=&-q^2g_{\mu\nu }+2(k_{1\mu }k_{2\nu }+k_{1\nu }k_{2\mu })+
2i\lambda _e<\mu\nu qk_2>, \nonumber\\
L_{\mu\nu }^{(i)}&=&-2i<\mu\nu k_1k_2>+\lambda _e[
q^2g_{\mu\nu }-2(k_{1\mu }k_{2\nu }+k_{1\nu }k_{2\mu })],
\label{eq:eq9}
\end{eqnarray}
where $<\mu\nu ab>=\varepsilon_{\mu\nu\rho\sigma }a_{\rho}b_{\sigma}$ and 
$\lambda _e$ is the degree of the electron longitudinal polarization.
The other components of the electron polarization lead to a  
suppression by a factor $m_e/m$, where $m_e$ is the electron mass. 

Taking into account the polarization states of the produced baryon and
antibaryon, the hadronic tensors can be written as a sum of three contributions:
\begin{equation}\label{eq:eq10}
H_{\mu\nu }^{(k)}=H_{\mu\nu }^{(k)}(0)+H_{\mu\nu }^{(k)}(1)+
H_{\mu\nu }^{(k)}(2), \ \ k=v, i,
\end{equation}
where the tensor $H_{\mu\nu }^{(k)}(0)$ describes the production of unpolarized 
particles, the tensor $H_{\mu\nu }^{(k)}(1)$ describes the production of 
polarized nucleon or antinucleon and the tensor $H_{\mu\nu }^{(k)}(2)$ corresponds 
to the production of both polarized particles, $N$ and $\bar N$.

Let us firstly consider the production of unpolarized $N\bar N-$ pair as a result of
annihilation of unpolarized $e^+e^--$ pair. In this case the general structure 
of the hadronic tensors can be written as 
\begin{equation}
H_{\mu\nu }^{(v)}(0)=H_1\tilde g_{\mu\nu }+H_2P_{\mu }P_{\nu }, \ 
H_{\mu\nu }^{(i)}(0)=-4iA_NG_{MN}^*<\mu\nu p_1p_2>, 
\label{eq:eq11}
\end{equation}
where $\tilde g_{\mu\nu }=g_{\mu\nu }-q_{\mu }q_{\nu}/q^2.$ One can get the
following expressions for these structure functions when the 
hadronic current is given by Eq. (\ref{eq:mat})
\begin{eqnarray}
H_1&=&-2q^2(|G_{MN}|^2+2ReG_{MN}\Delta G_{MN}^*), \label{eq:eq12} \\
H_2&=&\frac{8}{\tau -1}\biggl [|G_{EN}|^2-\tau |G_{MN}|^2+
2ReG_{EN}\Delta G_{EN}^*-2\tau ReG_{MN}\Delta G_{MN}^*\biggr ].\nonumber
\end{eqnarray}
The differential cross section of the reaction $ e^++e^-\rightarrow N+\bar N$, 
for the case of unpolarized 
particles, has the form 
\begin{equation}\label{eq:eq13}
\frac{d\sigma_{un}}{d\Omega } = \frac{\alpha^2\beta }{4q^2}D, \
\end{equation}
with
$$D=(1+\cos^2\theta )(|G_{MN}|^2+2ReG_{MN}\Delta G_{MN}^*)+\frac{1}{\tau }
\sin^2\theta (|G_{EN}|^2+ $$
$$+2ReG_{EN}\Delta G_{EN}^*)- 
\frac{4}{\tau }\sqrt{\tau (\tau -1)}\cos\theta Re G_{MN}A_N^*, $$
where $\theta $ is the angle between the momenta of the electron and the detected antinucleon,  
in the $e^++e^-\rightarrow N+\bar N$ reaction CMS. Note that Eq. (\ref{eq:eq13}) was 
obtained neglecting the terms of the order of $\alpha ^2$ compared to the 
dominant (Born approximation) terms. In the one--photon--exchange limit the 
expression  (\ref{eq:eq13}) coincides with the result obtained for the differential cross section in Ref. \cite{Du96}. The TPE contribution brings three new terms which are all 
of the order of $\alpha $ compared to the Born contribution. 

At the threshold of the reaction, $q^2 = 4m^2,$ the equality $G_{MN}=G_{EN}=G_N$ holds (this relation follows  from the definition (\ref{eq:eq4}))
and the  Eq. (\ref{eq:eq13})  reduces to 
$$\frac{d\sigma_{un}^{th}}{d\Omega } = \frac{\alpha^2\beta }{2q^2}D^{th}, $$
$$\ \ D^{th}=|G_{N}|^2+ReG_{N}(\Delta G_{MN}^*+\Delta G_{EN}^*)+
\cos^2\theta ReG_{N}(\Delta G_{MN}^*-\Delta G_{EN}^*). $$

As it was shown in Ref. \cite{Re04a}, symmetry properties of the amplitudes
with respect to the $\cos\theta \to -\cos\theta $ transformation can be derived from the $ C $  
invariance of the considered $1\gamma \otimes 2\gamma $ mechanism:
\begin{equation}
\Delta G_{MN,EN}(\cos\theta )=-\Delta G_{MN,EN}(-\cos\theta ),\ A_N(\cos\theta )=A_N(-\cos\theta ).
\label{eq:eq14}
\end{equation}
Let us consider the situation when the experimental apparatus does not distinguish the nucleon 
from the antinucleon. Then we measure the following sum of the differential cross sections
$$\frac{d\sigma_+}{d\Omega}=\frac{d\sigma}{d\Omega}(\cos\theta )+\frac{d\sigma}{d\Omega}(-\cos\theta ). $$
We can stress, using the properties (\ref{eq:eq14}), that this quantity does not depend on the TPE 
terms. This statement agrees with the conclusion of the paper \cite{Pu61}: for processes 
of the type $e^++e^-\to a^++a^-$, if the apparatus which detects the final particles does not distinguish 
the particle $a^+$ from the particle $a^-$, then the 
interference term between the matrix elements corresponding to the one--photon and two--photon 
exchange diagrams does not contribute to the cross section. 
 
Note also that the TPE terms do not contribute to the total cross section of the reaction 
$ e^++e^-\rightarrow N+\bar N$,  which can be written as 
\begin{equation}\label{eq:eq15}
\sigma _t(q^2)=\frac {4\pi }{3}\frac {\alpha ^2\beta }{q^2}\left [|G_{MN}(q^2)|^2+\frac{1}{2\tau }|G_{EN}(q^2)|^2
\right ]. 
\end{equation} 
On the other hand, the relative contribution of TPE mechanism is enhanced in the following  
angular asymmetry 
\begin{equation}\label{eq:eq16}
A(q^2,\theta_0)=\frac {\sigma (q^2,\theta_0)-\sigma (q^2,\pi -\theta_0)}
{\sigma (q^2,\theta_0)+\sigma (q^2,\pi -\theta_0)}, 
\end{equation}  
where the quantities $\sigma (q^2,\theta_0)$ and $\sigma (q^2,\pi -\theta_0)$ are defined as follows
$$\sigma (q^2,\theta_0)=\int_0^{\theta_0}\frac{d\sigma }{d\Omega}(q^2,\theta )d\Omega , \ \ 
\sigma (q^2,\pi -\theta_0)=\int_{\pi -\theta_0}^{\pi }\frac{d\sigma }{d\Omega}(q^2,\theta )d\Omega . $$ 
Using the symmetry relations (\ref{eq:eq14}) one  can obtain for the asymmetry $A(q^2, \theta_0)$ the  following expression
\begin{eqnarray}
A(q^2, \theta_0)&=&\frac{2}{d}\int_0^{\theta_0}dcos\theta \Bigg [(1+cos^2\theta )ReG_{MN}(q^2)\Delta G_{MN}^*(q^2,cos\theta )
\nonumber \\
&&+\frac{sin^2\theta }{\tau }ReG_{EN}(q^2)\Delta G_{EN}^*(q^2,cos\theta ) \nonumber \\
&&-\frac{2}{\tau }\sqrt{\tau (\tau -1)}cos\theta ReG_{MN}(q^2)A_N^*(q^2,cos\theta )\Bigg ],
\label{eq:eq17}
\end{eqnarray}
where the quantity $d$ is
$$d=\frac{1-x_0}{3}\Bigg [(4+x_0+x_0^2)|G_{MN}|^2+\frac{1}{\tau }(2-x_0-x_0^2)|G_{EN}|^2 \Bigg ], \ \ x_0=cos\theta_0 . $$

The TPE contributions can be removed considering the sum of the quantities $\sigma (q^2,\theta_0)$ 
and $\sigma (q^2,\pi -\theta_0)$. As a result we have
\begin{equation}
\Sigma (q^2, \theta_0)=\sigma (q^2,\theta_0)+\sigma (q^2,\pi -\theta_0)=\frac{\pi \alpha^2}{q^2}\beta d.
\label{eq:eq18}
\end{equation}
The terms of the order of $\alpha ^2$ are everywhere neglected.

%%%%%%%%%%%%%%%%%%%%%%%%%%%%%%%%%%%%%%%%%%%%%%%%%%%%%%%%%%%
\section{Single--spin polarization observables}
%%%%%%%%%%%%%%%%%%%%%%%%%%%%%%%%%%%%%%%%%%%%%%%%%%%%%%%%%%%%%%

Let us consider single--spin observables and calculate the hadronic 
tensors when the produced antinucleon is polarized. One finds
\begin{eqnarray}
H_{\mu\nu }^{(v)}(1)&=&\frac{2}{m}\frac{1}{\tau -1}\biggl [
im^2(\tau -1)|\tilde G_{MN}|^2<\mu\nu qs_1>+ \nonumber \\
&&
iRe\tilde G_{MN}(\tilde G_{EN}-\tilde G_{MN})^*(<\nu p_2p_1s_1>P_{\mu }- 
<\mu p_2p_1s_1>P_{\nu })+ \nonumber \\
&&Im\tilde G_{MN}\tilde G_{EN}^*(<\nu p_2p_1s_1>P_{\mu }+<\mu p_2p_1s_1>P_{\nu })\biggr ], \nonumber \\
H_{\mu\nu }^{(i)}(1)&=&mA_N\biggl [(\tilde G_{MN}+\tilde G_{EN})^*(s_{1\mu }p_{2\nu }+
s_{1\nu }p_{2\mu })+ \nonumber \\
&&(\tilde G_{MN}-\tilde G_{EN})^*(s_{1\mu }p_{1\nu }+s_{1\nu }p_{1\mu })-\nonumber \\
&& 
(\tilde G_{MN}+\tilde G_{EN})^*(s_{1\mu }p_{1\nu }-
s_{1\nu }p_{1\mu })- \label{eq:eq19} \\
&&(\tilde G_{MN}-\tilde G_{EN})^*(s_{1\mu }p_{2\nu }-s_{1\nu }p_{2\mu })-
\nonumber \\
&&
2q\cdot s_1\tilde G_{MN}^*g_{\mu\nu }-\frac{2}{m^2}\frac{q\cdot s_1}{1-\tau }(\tilde G_{MN}
-\tilde G_{EN})^*p_{1\mu }P_{\nu }\biggr ], \nonumber
\end{eqnarray}
where $s_{1\mu } \ (s_1\cdot p_1=0)$ is the antinucleon polarization four--vector. 

The polarization four--vector of a particle, of mass $m$ and energy $E$, in the system where its momentum is
${\vec p}$, is connected with the polarization vector ${\vec \chi}$ in its rest
frame by a Lorentz boost
$${\vec s}={\vec \chi}+\frac{{\vec p}\cdot {\vec \chi}{\vec p}}{m(E+p)}, \ 
s^0=\frac{1}{m}{\vec p}\cdot {\vec \chi }.$$

Let us define a coordinate frame in CMS of the reaction $ e^++e^-\rightarrow N+\bar N$ where the $z$ axis is
directed along the momentum of the antinucleon $({\vec p})$, the $y$ axis is 
orthogonal to the reaction plane and directed along the vector 
${\vec k}\times {\vec p}$, (${\vec k}$ is the electron momentum), 
and the $x$ axis forms a left--handed coordinate system. Therefore, the
components of the unit vectors are: $\hat {\vec p}=(0,0,1)$ and 
$\hat {\vec k}=(-\sin\theta ,0,\cos\theta)$ with 
$\hat {\vec p}\cdot \hat {\vec k}=\cos\theta.$

Note that, unlike elastic electron--nucleon scattering in the Born approximation, the hadronic 
tensor $H_{\mu\nu }^{(v)}(1)$ in the TL region contains a symmetric part 
even in the Born approximation due to the complexity of the nucleon FFs. 
Taking into account the TPE contribution leads to antisymmetric terms in 
the $H_{\mu\nu }^{(i)}(1)$ tensor. So, these terms lead to non--zero polarization 
of the outgoing antinucleon (the initial state is unpolarized), which can be written as 
\begin{eqnarray}
P_y&=&\frac{2sin\theta }{\sqrt{\tau }D}
\Bigg \{ cos\theta \left [
ImG_{MN}G_{EN}^*+Im(G_{MN}\Delta G_{EN}^*-G_{EN}\Delta G_{MN}^*)\right ]-\nonumber \\
&&-\sqrt{\frac{(\tau -1)}{\tau }}ImG_{EN}A_N^*\Bigg  \}.
\label{eq:eq20}
\end{eqnarray}
In the one--photon--exchange (Born) approximation this expression gives the
well known result for the polarization $P_y$ obtained in Ref. \cite{Du96}.
One can see also that

- The polarization of the outgoing antinucleon in this case is determined
by the polarization component which is perpendicular to the reaction plane.

- The polarization, being T--odd quantity, does not vanish even in the
one--photon--exchange approximation due to the complexity of the nucleon FFs
in the TL region (to say more exactly, due to the non--zero difference
of the phases of these FFs). This is principal difference with the elastic
electron--nucleon scattering.

- In the Born approximation this polarization becomes equal to zero at the
scattering angle $\theta = 90^0$ (as well at $\theta = 0^0$ and $180^0$).
The presence of the TPE contributions leads to a non--zero value of the
polarization at this angle and it is determined by a simple expression
$$P_y(90^0)=-2\frac{\sqrt{\tau -1}}{\tau\bar D}ImG_{EN}A_N^*,
\ \ \bar D=D(\theta =90^0). $$
Here the function $A_N$ is also taken at the value $\theta =90^0$.
This quantity expected to be small due to the fact that it is determined by
the interference of the one--photon and two--photon exchange amplitudes and
may be of the order of $\alpha $. The measurement of this polarization at
$\theta = 90^0$ can give information about the TPE contribution and its
behaviour as a function of $q^2.$

In the threshold region we can conclude that in the Born approximation this polarization vanishes, 
due to the relation $G_{EN}=G_{MN}$ which is valid at the threshold. The TPE contributions induces a non zero polarization, which is determined by a simple formula
$$P_y^{th}(\theta )=\frac{sin2\theta }{D^{th}}ImG_N(
\Delta G_{EN}-\Delta G_{MN})^*. $$
Note that, at threshold, this polarization can still vanish if
$\Delta G_{EN}=\Delta G_{MN}$. In this case the differential cross section does not contain any explicit dependence on the angular variable $\theta $.
In the general case, the amplitudes $\Delta G_{EN,MN}$ depend on the $\theta $
variable. The effect of the TPE contributions for the polarization at an arbitrary scattering angle is expected to increase as $q^2$ increase, as the TPE amplitudes decrease more slowly with $q^2$ in comparison with the nucleon FFs.

Using the properties of the TPE amplitudes with respect to the 
$\cos\theta \to -\cos\theta $ transformation, we can remove the contributions of
the TPE effects by constructing the following quantities. Let us introduce the terms
$P_y(q^2,\theta_0)$ and $P_y(q^2,\pi -\theta_0)$, which are integrals of the 
polarization $P_y(q^2,\theta )$ over the angular regions connected by 
the above  mentioned transformation
$$P_y(q^2,\theta_0)=\int_0^{\theta_0}P_y(q^2,\theta )d\Omega , ~P_y(q^2,\pi -\theta_0)=\int_{\pi -\theta_0}^{\pi }P_y(q^2,\theta )d\Omega . $$ 
Let us calculate the sum and the difference of these two quantities. At first order of the coupling constant $\alpha $, we obtain
\begin{eqnarray}
&&D^P(q^2,\theta_0)=P_y(q^2,\theta_0)-P_y(q^2,\pi -\theta_0)=\nonumber \\
&&=-\frac{8\pi R}{\sqrt{\tau }}(1-\frac{R^2}{\tau })^{-(3/2)}
\sin(\delta_{MN}-\delta_{EN})\Bigg
[\sqrt{z}+\frac{1}{\sqrt{2}}ln \left |\frac{\sqrt{z}-\sqrt{2}}{\sqrt{z}+\sqrt{2}}\right |\Bigg ], 
\label{eq:eq21}
\end{eqnarray}
where 
$$R=\frac{|G_{EN}|}{|G_{MN}|}, \ \ z=(1-x_0^2)(1-\frac{R^2}{\tau }), $$ 
and $\delta_{MN}(\delta_{EN})$  is the phase of the complex FF $G_{MN}(G_{EN})$. We can 
see that, in this approximation, the quantity $D^P$ does not depend on the TPE 
contribution. So, the phase difference of FFs can be correctly determined from this quantity, if the ratio $R$ is known. 

Let us consider the ratio of the function $\Sigma (q^2,\theta_0)$, Eq. (\ref{eq:eq18}), calculated at two values of $\theta_0$: 
$$\frac{\Sigma (q^2,\theta_1)}{\Sigma (q^2,\theta_2)}=\frac{1-x_1}{1-x_2}\cdot  
 \frac{4+x_1+x_1^2+\frac{1}{\tau }(2-x_1-x_1^2)R^2}{4+x_2+x_2^2+\frac{1}{\tau }(2-x_2-x_2^2)R^2}, 
 \ \  x_i=\cos\theta_i, \ i=1,2 $$
This ratio allows to determine $R$, minimizing systematic errors.

The magnitude of the TPE contribution to the polarization $P_y$, integrated over the considered angular region, can be obtained from the sum of the quantities
 introduced above
\begin{eqnarray}
&\Sigma ^P(q^2,\theta_0)=&P_y(q^2,\theta_0)+P_y(q^2,\pi -\theta_0)=\nonumber \\
&=&\frac{8\pi }{\sqrt{\tau }}\int_0^{\theta_0}d\cos\theta \frac{sin\theta }{D_B}\Bigg 
\{\cos\theta Im(G_{MN}\Delta G^*_{EN}-G_{EN}\Delta G^*_{MN})- \nonumber \\
&&2\frac{\cos\theta }{D_B}ImG_{MN}G^*_{EN}\Bigg [(1+\cos^2\theta )ReG_{MN}\Delta G^*_{MN}+\label{eq:eq22}\\
&&
\frac{\sin^2\theta }{\tau }ReG_{EN}\Delta G^*_{EN}\Bigg ]- \nonumber \\
&&\sqrt{\frac{\tau -1}{\tau }}\Bigg [ImG_{EN}A_N^*+4\frac{\cos^2\theta }{D_B}
ImG_{MN}G^*_{EN}ReG_{MN}A^*_{N}\Bigg ]\Bigg \}, \nonumber
\end{eqnarray} 
where  
$$D_B=(1+\cos^2\theta )|G_{MN}|^2+\frac{\sin^2\theta }{\tau }|G_{EN}|^2. $$
Let us consider the single--spin asymmetry induced by the transverse polarization of the 
electron or positron beam. The leptonic tensors corresponding to an arbitrary polarized electron beam have the form
\begin{eqnarray}
L_{\mu\nu }^{(v)}(s_e)&=&2im_e<\mu\nu s_eq>, \nonumber\\
L_{\mu\nu }^{(i)}(s_e)&=&-2m_e[-k_2\cdot s_eg_{\mu\nu }+k_{2\mu }s_{e\nu }+k_{2\nu }s_{e\mu }+
k_{1\mu }s_{e\nu }-k_{1\nu }s_{e\mu }],
\label{eq:eq23}
\end{eqnarray}
where $s_{e\mu} \ (s_e\cdot k_1=0)$ is the electron polarization four--vector.

Then the asymmetry can be written as
\begin{eqnarray}
A_e=4\frac{m_e}{\sqrt{q^2}}\frac{\beta}{D_B}ImG_{MN}A_N^*(\hat{\vec k} \times\hat{\vec p})\cdot \vec\xi ,
\label{eq:eq24}
\end{eqnarray} 
where ${\vec \xi}$ is the unit vector along the polarization of the electron in its rest system. $A_e$ is determined by the electron spin vector component which is perpendicular to the reaction plane. It vanishes in the Born approximation as it is determined by the imaginary 
part of the product $G_{MN}A_N^*$, i.e., by the spin structure induced by the TPE contribution.
The same term defines the magnitude of the $P_{xy}$ and $P_{yx}$ components of the polarization 
correlation tensor of the nucleon and antinucleon. Moreover, $A_e$ is proportional to the electron mass and vanishes for scattering angles $\theta =0^0$ and $180^0.$

%%%%%%%%%%%%%%%%%%%%%%%%%%%%%%%%%%%%%%%%%%%%%%%%%%%%%%%%%%%
\section{Double--spin polarization observables}
%%%%%%%%%%%%%%%%%%%%%%%%%%%%%%%%%%%%%%%%%%%%%%%%%%%%%%%%%%%
If one of the colliding beam is longitudinally polarized then the antinucleon acquires 
$x-$ and $z-$components of the polarization, which lie in the 
$e^+e^-\to N\bar N$ reaction plane. These components can be written as (we 
assume 100 \% polarization of the electron beam)
\begin{eqnarray}
P_x&=&-\frac{2\sin\theta }{\sqrt{\tau }D}\biggl [Re(G_{MN}G_{EN}^*+
G_{MN}\Delta G_{EN}^*+G_{EN}\Delta G_{MN}^*)-
\nonumber \\
&&-\sqrt{\frac{\tau -1}{\tau }}\cos\theta ReG_{EN}A_N^*\biggr],\label{eq:eq25} \\
P_z&=&\frac{2}{D}\biggl [\cos\theta (|G_{MN}|^2+2ReG_{MN}\Delta G_{MN}^*)+
\sqrt{\frac{\tau -1}{\tau }}\sin^2\theta ReG_{EN}A_N^*\biggr].  \nonumber
\end{eqnarray}
These polarization components are T--even observables and they are non--zero
also for the elastic electron--nucleon scattering in the Born approximation.
Note that in  the Born approximation we obtained the result of Ref. \cite{Du96}. 

Transversally polarized electron beams lead to a 
polarization for the antinucleon, which is a factor $(m_e/m)$ smaller than for the case of the
longitudinal polarization of the electron beam.

The polarization component $P_z$ vanishes when the proton is emitted at an 
angle $\theta = 90^0$ in the Born approximation. But the presence of the TPE
term $A_N$ in the electromagnetic hadron current may lead to non--zero value 
of this quantity if the amplitude $A_N(\theta = 90^0)$ is not zero, since the 
value of this component is determined by the term $ReG_{EN}A_N^*.$

Let us consider the case when the produced antinucleon and nucleon are
both polarized. The corresponding hadronic tensors can be written as
\begin{eqnarray}
H_{\mu\nu }^{(v)}(2)&=&C_1g_{\mu\nu }+C_2P_{\mu }P_{\nu }+
C_3(P_{\mu }s_{1\nu }+P_{\nu }s_{1\mu})+C_4(P_{\mu }s_{2\nu }+
P_{\nu }s_{2\mu})+ \nonumber \\
&& C_5(s_{1\mu }s_{2\nu }+s_{1\nu }s_{2\mu})+
iC_6(P_{\mu }s_{1\nu }-P_{\nu }s_{1\mu})+ 
iC_7(P_{\mu }s_{2\nu }-P_{\nu }s_{2\mu}),
\nonumber \\
H_{\mu\nu }^{(i)}(2)&=&iA_N\Bigg [\frac{2}{\tau -1}(G_{MN}-G_{EN})^*P_{\nu}<\mu s_1s_2P>+
 \nonumber \\
&&
G_{MN}^*(m^2<\mu\nu s_1s_2>+q\cdot s_1<\mu\nu s_2p_1>+q\cdot s_2<\mu\nu s_1p_2>- \nonumber \\
&&
p_{1\mu}<\nu s_1s_2p_2>-
p_{2\nu}<\mu s_1s_2p_1>-
\nonumber \\
&&
s_{2\mu}<\nu p_2s_1p_1>-s_{1\nu}<\mu p_2s_2p_1>)\Bigg ], 
\label{eq:eq26}
\end{eqnarray}
where $s_{2\mu }$ is the nucleon polarization 4-vector ($p_2\cdot s_2 = 0$). 
We omitted the terms proportional to $q_{\mu }$ or $q_{\nu },$ since they
do not contribute to the cross section and to the polarization observables due to the
conservation of the leptonic current. The antisymmetrical part of the tensor
$H_{\mu\nu }^{(v)}(2)$ arises from the fact that nucleon FFs in the TL region 
are complex quantities.

The structure functions $C_i$ have the following form
\begin{eqnarray}
C_1&=&\frac{1}{2}(q^2s_1\cdot s_2-2q\cdot s_1q\cdot s_2)|\tilde G_{MN}|^2, \nonumber \\
C_2&=&\frac{2}{(\tau -1)^2}\left .[|\tau\tilde G_{MN}-\tilde G_{EN}|^2
s_1\cdot s_2+ \right .\nonumber \\
&&\left .\frac{1}{2m^2}(2q\cdot s_1q\cdot s_2-q^2s_1\cdot s_2)
|\tilde G_{EN}-\tilde G_{MN}|^2\right ], \nonumber \\
C_3&=&ReE_1, \ \  C_4=ReE_2,~
C_5=-\frac{q^2}{2}|\tilde G_{MN}|^2,\nonumber \\
E_1&=&\frac{q\cdot s_2}{\tau -1}(\tau |\tilde G_{MN}|^2-
\tilde G_{EN}\tilde G_{MN}^*), \ \
E_2=-\frac{q\cdot s_1}{\tau -1}(\tau |\tilde G_{MN}|^2-
\tilde G_{EN}\tilde G_{MN}^*), \nonumber \\
C_6&=&ImE_1, \ \  C_7=ImE_2.
\label{eq:eq27}
\end{eqnarray}
It follows that the components of the polarization correlation tensor $P_{ik}$, $(i,k=x,y,z)$, of the nucleon and the antinucleon, in presence of the one--photon--exchange plus
two--photon--exchange mechanisms in the $e^++e^-\to N+\bar N$ process, have the following expressions:
\begin{eqnarray}
P_{xx}&=&\frac{sin^2\theta }{\tau D}\biggl [\tau (|G_{MN}|^2+
2ReG_{MN}\Delta G_{MN}^*)+|G_{EN}|^2+2ReG_{EN}\Delta G_{EN}^*\biggr ],\nonumber \\
P_{yy}&=&\frac{sin^2\theta }{\tau D}\biggl [|G_{EN}|^2+2ReG_{EN}\Delta
G_{EN}^*-\tau (|G_{MN}|^2+2ReG_{MN}\Delta G_{MN}^*)\biggr ], \nonumber \\
P_{zz}&=&\frac{1}{\tau D}\biggl [\tau (1+\cos^2\theta )(|G_{MN}|^2+
2ReG_{MN}\Delta G_{MN}^*)-\sin^2\theta (|G_{EN}|^2+ \nonumber \\
&&2ReG_{EN}\Delta G_{EN}^*)-
4\sqrt{\tau (\tau -1)}\cos\theta ReG_{MN}A_N^*\biggr ], \nonumber \\
P_{xz}&=&P_{zx}=-2\frac{\sin\theta }{\sqrt{\tau }D}\biggl [\cos\theta
Re(G_{MN}G_{EN}^*+G_{MN}\Delta G_{EN}^*+G_{EN}\Delta G_{MN}^*)-\nonumber \\
&&
\sqrt{\frac{\tau -1}{\tau }}ReG_{EN}A_N^*\biggr ]. 
\label{eq:eq28}
\end{eqnarray}
For completeness we give also  the non--zero coefficients in case
of longitudinally polarized electron beam
\begin{eqnarray}
P_{xy}&=&P_{yx}=-\frac{1}{D}\sqrt{\frac{\tau -1}{\tau}}sin^2\theta
ImG_{MN}A_N^*,\nonumber \\
&&P_{zy}=P_{yz}=\frac{sin\theta }{\sqrt{\tau }D}Im(G_{MN}G_{EN}^*+
G_{MN}\Delta G_{EN}^*-G_{EN}\Delta G_{MN}^*+\nonumber \\
&&
+\sqrt{\frac{\tau -1}{\tau}}\cos\theta G_{EN}A_N^*). 
\label{eq:eq29}
\end{eqnarray}
One can easily verify that the following relation holds:
$$P_{xx}+P_{yy}+P_{zz}=1. $$

Let us note the following properties for these coefficients.

- The components of the tensor describing the polarization correlations
$P_{xx}, $ $P_{yy}, $ $P_{zz},$ $P_{xz},$ and $P_{zx}$ are T--even
observables, whereas the components $P_{xy}, $ $P_{yx}, $ $P_{yz},$ and
$P_{zy}$ are T--odd ones.

- In the Born approximation the expressions for the T--odd polarization
correlations coincide with the corresponding components of the
polarization correlation tensor of baryon $B$ and antibaryon $\bar B$
created by the one--photon--exchange mechanism in the $e^+e^-\to B\bar B$
process \cite{Du96} \footnote{The expressions in Eq. (24) of Ref. \protect\cite{Du96} should be: 
\begin{eqnarray*}
P_{xx}&=&\frac{sin^2\theta }{\tau D} [\tau |G_{MN}|^2+|G_{EN}|^2
], \nonumber \\
P_{yy}&=&\frac{sin^2\theta }{\tau D}
 [|G_{EN}|^2-\tau |G_{MN}|^2], \nonumber \\
P_{zz}&=&\frac{1}{\tau D }
[\tau (1+\cos^2\theta )|G_{MN}|^2-\sin^2\theta |G_{EN}|^2],  \nonumber \\
P_{xz}&=&P_{zx}=-2\frac{\sin 2\theta }{\sqrt{\tau }D} Re[G_{MN}G_{EN}^*],
\nonumber \\
P_{xy}&=&P_{yx}=0,\nonumber \\
P_{zy}&=&P_{yz}=\frac{sin\theta }{\sqrt{\tau }D}Im(G_{MN}G_{EN}^*).\nonumber 
\end{eqnarray*}
}.

- The relative contribution of the interference terms (between one- and
two--photon--exchange terms) in these observables will increase as
the value $q^2$ becomes larger since it is expected that the TPE amplitudes
decrease more slowly with $q^2$ compared with the nucleon FFs.

At the reaction threshold, the polarization correlation tensor components have
some specific properties:

- All correlation coefficients (both T--odd and T--even) do not depend on
the function $A_N$.

- In the Born approximation the $P_{yy}$ observable is zero, but the presence 
of the TPE term leads to a non--zero value, determined by the quantity
$ReG_{N}(\Delta G_{EN}-\Delta G_{MN})^*.$

- At the scattering angle $\theta = 90^0$ we have the relation $P_{yy}+
P_{zz}=0.$

- The $P_{xy}$ and $P_{yx}$ observables are zero, and $P_{yz}$ and $P_{zy}$
observables are determined by the TPE term only, namely by the quantity
$ImG_{MN}(\Delta G_{EN}-\Delta G_{MN})^*.$

%%%%%%%%%%%%%%%%%%%%%%%%%%%%%%%
\section{Conclusions}
%%%%%%%%%%%%%%%%%%%%%%%%%%%%%%%

Precise measurements of the elastic electron--hadron scattering arised the question of the importance of the 
TPE mechanism. This problem enters also in the determination of the nucleon 
FFs in the TL region investigated with the $e^++e^-\to N+\bar N$ reactions, since this process is a crossed channel of the elastic electron--nucleon scattering. 

The influence of the TPE contribution on the observables of the $e^++e^-\to N+\bar N$
reaction was investigated in detail starting from a general parametrization of the relevant matrix element. The expressions for the differential cross section and 
various polarization observables (such as the antinucleon polarization, the polarization 
correlation coefficients of the produced $N\bar N$ pair, the polarization of the antinucleon due to the electron longitudinal polarization, and the single--spin asymmetry 
caused by the transverse polarization of the electron beam) were derived in terms of 
three complex amplitudes.

Using the properties of these amplitudes with respect to the $\cos\theta \to -\cos\theta $
transformation, it was shown how one can remove or enhance the TPE contribution from various 
observables. The TPE effects have been singled out in particular observables, when possible.

We showed that the real part of the TPE 
matrix element can be accesssed through the difference between the differential cross sections 
measured at angles $\theta $ and $\pi -\theta $, whereas for the imaginary part of the TPE matrix element one needs the measurement of the single--spin observables (the asymmetry caused 
by a transversely polarized electron beam or by the nucleon or antinucleon polarization when 
all the other particles are unpolarized).

Single--spin observables such as the polarization of the produced proton in $e^+e^-$ 
annihilation may be more difficult to measure than in the electron--proton elastic scattering, due 
to luminosity. On the other hand, angular distribution is easier to achieve as it is necessary to 
make measurements at different angles, with fixed beam energy, whereas in scattering reactions 
measurements should be done at fixed momentum transfer squared, which requires to change 
simultaneously beam energy and scattering angle. 

The present analysis of the TPE effects in the $e^++e^-\to N+\bar N$ reaction  should be helpful for the experimental 
investigation of the TL nucleon FFs planned in near future at Frascati and Novosibirsk. 

These results complete our previous work on model independent derivation of polarization 
observables in the electron--proton elastic scattering and in the crossing channels, 
$e^++e^-\to N+\bar N$ and  $N+\bar N\to e^++e^-$.

\section{Acknowledgments}
The authors are grateful to M. P. Rekalo for enlightning discussions and essential contributions to the two--photon exchange problem.

Thanks are due to the organizers and to the participants of the N05 Workshop, hold on 12-14 October 2005, in Frascati, for exchanges and discussions very helpful for finalizing the present work.


\begin{thebibliography}{99}
\bibitem{Ho62} %\cite{Hofstadter:ae}
R.~Hofstadter, F Bumiller and M. Yearian,
%``Electron Scattering From The Proton,''
 Rev.\ Mod. Phys. \ 30 (1958) 482.
%%CITATION = PHRVA,98,217;%%
\bibitem{Ro50} M. N. Rosenbluth, Phys. Rev. 79 (1950) 615.
\bibitem{Re68} A. Akhiezer and M. P. Rekalo, Dokl. Akad. Nauk USSR 180 (1968) 1081; 
Sov. J. Part. Nucl.  4 (1974) 277.

\bibitem{JG00}
M. K. Jones {\it et al.}, Phys. Rev. Lett. 84 (2000) 1398;
O. Gayou {\it et al.}, Phys. Rev. Lett. 88 (2002) 092301;
%\cite{Punjabi:2005wq}
  V.~Punjabi {\it et al.},
  %``Proton elastic form factor ratios to Q**2 = 3.5-GeV**2 by polarization
  %transfer,''
  Phys.\ Rev.\ C 71 (2005) 055202.
%  [Erratum-ibid.\ C {\bf 71} (2005) 069902]
%  [arXiv:nucl-ex/0501018].
  %%CITATION = NUCL-EX 0501018;%%
\bibitem{Ar05}
%\cite{Qattan:2004ht}
I.~A.~Qattan {\it et al.},
%``Precision Rosenbluth measurement of the proton elastic form factors,''
Phys. Rev. Lett. 94 (2005) 142301.
%%CITATION = NUCL-EX 0410010;%%
\bibitem{Gu03}
P.A.M. Guichon and M. Vanderhaeghen,
Phys. Rev. Lett.  91 (2003) 142303.
\bibitem{Bl03}
P. G. Blunden, W. Melnitchouk, and J. A. Tjon,
Phys. Rev. Lett. 91 (2003) 142304.
\bibitem{Ch04}
Y.-C. Chen, A. Afanasev, S. J. Brodsky, C. E. Carlson, and  M. Vanderhaeghen,
Phys. Rev. Lett.  93 (2004) 122301.
\bibitem{Re03}
M.~P.~Rekalo and E.~Tomasi-Gustafsson,
%``Model independent properties of two-photon exchange in elastic electron
%proton scattering,''
Eur.  Phys. J.  A22 (2004) 331.
%%CITATION = NUCL-TH 0307066;%%
\bibitem{ETG05}
E.~Tomasi-Gustafsson and G.~I.~Gakh,
%``Search for evidence of two photon contribution in elastic electron proton
%data,''
Phys. Rev. C 72 (2005) 015209. 

\bibitem{Re99} M. P. Rekalo, E. Tomasi-Gustafsson and D. Prout, 
Phys. Rev. C 60 (1999)  042202.

\bibitem{Al99} L. C. Alexa {\it et al.}, Phys. Rev. Lett. 82 (1999) 1374.
\bibitem{Ab99} D. Abbott {\it et al.}, Phys. Rev. Lett. 82 (1999) 1379.

\bibitem{Gu73}
J. Gunion, L. Stodolsky,
Phys. Rev. Lett. 30 (1973) 345.
\bibitem{Fr73}
V. Franco, Phys. Rev. D8 (1973) 826.
\bibitem{Bo73}
V. N. Boitsov, L.A. Kondratyuk, and V. B. Kopeliovich,
Sov. J. Nucl. Phys. 16 (1973) 238;
F. M. Lev, Sov. J. Nucl. Phys. 21 (1973) 45.
%%CITATION = HEP-PH 0412137;%%
\bibitem{An03}
M. Andreotti {\it et al.},
Phys. Lett. B559 (2003) 20.

\bibitem{Falco}
A. De Falco,
Workshop on "$e^+e^-$ in the 1-2GeV range: Physics and Accelerator Prospects", 10-13 Sept. 2003,
Alghero (Italy); hep-ex/0312012.
\bibitem{An98}
A. Antonelli {\it et al.},
Nucl. Phys. B517 (1998) 3.
\bibitem{Ba05} R. Baldini [for the BaBar collaboration] E.P.S. Conference, Lisbon, July 21-27, (2005) and refs herein.

\bibitem{Ba94}
B. Bardin {\it et al.},
Nucl. Phys. B 411 (1994) 3.

\bibitem{PANDA}
An International Accelerator Facility for Beams of Ions and Antiprotons, 
{\it Conceptual Design Report}, http://www.gsi.de

\bibitem{PAX}
%\cite{Lenisa:2005pu}
  P.~Lenisa and F.~Rathmann  [the PAX Collaboration],
  %``Antiproton proton scattering experiments with polarization,''
  arXiv:hep-ex/0505054.
  %%CITATION = HEP-EX 0505054;%%

\bibitem{Fi03}
A. Filippi 
%"Precision measurements in the nucleon form factor in the time--like
%region with $FINUDA$ at $DA\Phi NE2$",
Workshop on "$e^+e^-$ in the 1-2GeV range: Physics and Accelerator Prospects", 10-13 Sept. 2003,
Alghero (Italy).

\bibitem{Du96}
A. Z. Dubnickova, S. Dubnicka and M. P. Rekalo,
Nuovo Cim. A109 (1996) 241.

\bibitem{Br04}
S. J. Brodsky, C. E. Carlson, J. R. Hiller and D. S. Hwang,
Phys. Rev. D69 (2004) 054022.

\bibitem{ETG05a} 
%\bibitem{Tomasi-Gustafsson:2005kc}
  E.~Tomasi-Gustafsson, F.~Lacroix, C.~Duterte and G.~I.~Gakh,
  %``Nucleon electromagnetic form factors and polarization observables in
  %space-like and time-like regions,''
  Eur.\ Phys.\ J.\ A  24 (2005) 419.
%  [arXiv:nucl-th/0503001].
  %%CITATION = NUCL-TH 0503001;%%
  
\bibitem{Re04}
M. P. Rekalo, E. Tomasi-Gustafsson,
Nucl. Phys. A740 (2004) 271.
\bibitem{Re04a}
M. P. Rekalo, E. Tomasi-Gustafsson,
Nucl. Phys. A742 (2004) 322.

\bibitem{Ga05}
G.~I.~Gakh, E.~Tomasi-Gustafsson,   
  %``Nucleon electromagnetic form factors and polarization observables in
  %space-like and time-like regions,''
  Nucl.\ Phys. A761 (2005) 120.
%  [arXiv:nucl-th/0503001].
  %%CITATION = NUCL-TH 0503001;%


\bibitem{Ni05} D. Nikolenko, private communication.

\bibitem{Fl63}
D. Flamm and W. Kummer, Nuovo Cim. 28 (1963) 33.
S. D. Drell and J. D. Sullivan, Phys. Lett. 19 (1965) 516. 
\bibitem{Pu61}
G. Putzolu, Nuovo Cim. 20 (1961) 542.
\end{thebibliography}
\end{document}